\documentclass[11pt,a4paper]{article}
\pdfoutput=1
\usepackage{graphicx}
\usepackage{color}
\usepackage{jcappub}
\usepackage{amsmath} 
   \usepackage{multirow}

\newcommand{\be}{\begin{equation}}
\newcommand{\ee}{\end{equation}}
\newcommand{\bea}{\begin{eqnarray}}
\newcommand{\eea}{\end{eqnarray}}

\newcommand{\figref}[1]{Fig.~\ref{fig:#1}}

\newcommand{\gsim}{\lower.7ex\hbox{$\;\stackrel{\textstyle>}{\sim}\;$}}
\newcommand{\lsim}{\lower.7ex\hbox{$\;\stackrel{\textstyle<}{\sim}\;$}}

\title{
Uncertainties in Direct Dark Matter Detection in Light of Gaia's Escape Velocity Measurements
}

\author[a]{Youjia Wu,}

\author[a,b,c]{Katherine Freese,}

\author[d]{Chris Kelso,}

\author[b]{Patrick Stengel}

\author[e]{and Monica Valluri}

\affiliation[a]{Leinweber Center for Theoretical Physics, Department of Physics, University of Michigan, Ann Arbor, MI 48109, USA}
\affiliation[b]{The Oskar Klein Centre for Cosmoparticle Physics,
	Department of Physics,
	Stockholm University,
	AlbaNova,
	10691 Stockholm,
	Sweden}
\affiliation[c]{Nordita,
	KTH Royal Institute of Technology and Stockholm University
	Roslagstullsbacken 23,
	10691 Stockholm,
	Sweden}
\affiliation[d]{Department of Physics, University of North Florida, Jacksonville, FL  32224, USA}
\affiliation[e]{Department of Astronomy, University of Michigan, Ann Arbor, MI 48109, USA}

\abstract{
Direct detection experiments have set increasingly stringent limits on the cross section for spin-independent dark matter-nucleon interactions. In obtaining such limits, experiments primarily assume the standard halo model (SHM) as the distribution of dark matter in our Milky Way. Three astrophysical parameters are required to define the SHM: the local dark matter escape velocity, the local dark matter density and the circular velocity of the sun around the center of the galaxy. This paper studies the effect of the uncertainties in these three astrophysical parameters on the XENON1T exclusion limits using the publicly available DDCalc code. We compare limits obtained using the widely assumed escape velocity from the RAVE survey and the newly calculated escape velocity by Monari $et~al.$ using Gaia data. Our study finds that the astrophysical uncertainties are dominated by the uncertainty in the escape velocity (independent of the best fit value) at dark matter masses below 6~GeV and can lead to a variation of nearly 6 orders of magnitude in the exclusion limits at 4 GeV.  Above a WIMP mass of 6~GeV, the uncertainty becomes dominated by the local dark matter density, leading to uncertainties of factors of $\sim$10 (3) at 6 (15) GeV WIMP mass in the exclusion limits.  Additionally, this work finds that the updated best fit value for the escape velocity based on Gaia data leads to only very minor changes to the effects of the astrophysical uncertainties on the XENON1T exclusion limits.}

\arxivnumber{1904.04781}
\notoc
\begin{document}

\maketitle

\section{Introduction}

While the evidence of gravitational interactions due to Dark Matter (DM) is well established at a wide variety of scales and epochs in the universe, observation of DM interactions with the Standard Model (SM) remains elusive. One of the most well-studied and theoretically well-motivated DM candidates is the Weakly Interacting Massive Particle (WIMP). Signatures of non-gravitational WIMP interactions can be searched for through missing energy searches at LHC, by looking for the SM byproducts of WIMP annihilation in the Universe, and in Direct Detection (DD) experiments. 

Direct Detection experiments search for signatures of DM interactions with nuclei~\cite{Goodman:1984dc,Drukier:1986tm}. A significant experimental program is underway to search for nuclear recoils arising from the elastic scattering of WIMPs in the local DM halo. DD experiments are instrumented with various technologies designed to detect recoils of nuclei or electrons with kinetic energies as low as a few eV. In using experimental results to learn about dark matter properties, assumptions must be made about the local DM distribution. In this paper, we discuss the effect of several large astrophysical uncertainties in the DM distribution on the limits which have been set on the spin-independent (SI) WIMP-nucleon cross section.

Results from direct detection experiments are often analyzed assuming a simple isothermal distribution of dark matter, the Standard Halo Model (SHM). Although the SHM has been shown to be in tension with both cosmological simulations of galaxy formation and observations of baryonic matter in dwarf galaxies within the local group, the SHM is a reasonable first approximation for the calculation of the expected flux of DM particles at direct detection experiments~\cite{Bozorgnia:2016fjt,Kelso:2016qqj,Sloane:2016kyi}. The effects of variations in astrophysical parameters on the sensitivity of DD experiments, both within the context of and assuming various deviations from the SHM, have been investigated (for example, see~\cite{Green:2002ht,McCabe:2010zh,Fairbairn:2012zs,Benito:2016kyp,Green:2017odb,Fowlie:2017ufs,Ibarra:2018yxq,Fowlie:2018svr,Petac:2018wnu}). Previous work demonstrates that variations in the escape velocity and circular velocity of dark matter could yield a shift to exclusion limits at smaller WIMP masses in DD experiments (for example, see~\cite{McCabe:2010zh}). Interesting recent work~\cite{Necib:2018iwb} argues for the existence of local tidal debris which would modify the local distribution. These results rely on the argument that low metallicity stars serve as a proxy for dark matter, an idea that is still a matter of debate, as discussed in, for example, Ref.~\cite{Evans:2018bqy}. Ref.~\cite{Evans:2018bqy} also suggests a refinement to SHM to include the Gaia sausage, updates values of some astrophysical parameters in SHM, and then calculates the effects of these changes on the limits set by DD experiments. 
In the current paper, we restrict our studies to the SHM, as this is the primary distribution used by DD experiments for the interpretation of their data, and investigate the effect on dark matter exclusion limits of uncertainties in the astrophysical parameters used to define the SHM .

We focus on the effects of three astrophysical parameters and their uncertainties on dark matter DD studies:  the local escape velocity, the local circular velocity of the solar system around the galaxy and the local dark matter density. A variety of local and global observations of the Milky Way suggest a wide range of possible values for the local DM density, $\rho_{\chi}$~\cite{Salucci:2010qr,Catena:2009mf,Weber:2010,Iocco:2011jz,McMillan:2011,Pato:2015dua,McKee:2015hwa,Xia:2015agz,Huang:2016,McMillan:2016,Sivertsson:2017rkp,Buch:2018qdr,Widmark:2018ylf,deSalas:2019}. Although the recoil rate at DD experiments is simply $\propto \rho_{\chi}$, the uncertainty in the sensitivity of such experiments due to other astrophysical parameters is exacerbated by the possibly large variations in the local DM density. Especially at low WIMP masses, ${m_\chi} \lsim 15 \mathrm{GeV}$, nuclear recoils with enough kinetic energy to be detected above threshold at DD experiments must be induced by WIMPs at the high velocity tail of the local phase space distribution. Thus, uncertainties both in the shape of the local velocity distribution, characterized by the local circular velocity, and in the local escape velocity can significantly impact the sensitivity of DD experiments to low mass WIMPs.

In this paper, we investigate the impact of astrophysical uncertainties on the sensitivity of current DD experiments. Specifically, we use DDCalc software~\cite{Workgroup:2017lvb} to perform a likelihood analysis of the sensitivity of XENON1T~\cite{Aprile:2017iyp} to low mass WIMPs while sampling values of the relevant astrophysical parameters from the latest observational data. For the local escape velocity, we compare the sensitivity of DD experiments using the results of the RAVE survey~\cite{Piffl:2013mla} and a recent analysis of Gaia-DR2~\cite{Monari:2018}. We show that, for very low WIMP masses which would only produce recoils near the threshold of DD experiments, the sensitivity can vary by several orders of magnitude in the WIMP-nucleon cross section assuming either measurement of the escape velocity. Thus, a more precise determination of the escape velocity from data could significantly reduce the astrophysical uncertainties in DD experiments. Also, we find that the astrophysical uncertainties in the XENON1T exclusion limits are similar when assuming either the results of the RAVE survey or the more recent determination of the escape velocity based on Gaia data. 

This paper is structured as follows: in Section~\ref{sec:AstroPara} we discuss the relevant astrophysical parameters and their uncertainties, in Section~\ref{sec:Results} we present the results of the likelihood analyses and in Section~\ref{sec:Conclusions} we discuss our conclusions.

\section{Astrophysical Parameters}
\label{sec:AstroPara}
The differential recoil rate for the elastic scattering of target nuclei in DD experiments (per unit target mass) can be written as a function of the momentum transfer between the WIMP and the nucleus $q$, and WIMP velocity $v$, in the lab frame
\bea \label{eq:diffRecoilRate}
		\frac{dR}{dE_R} = 2 \frac{ \rho_{ \chi}}{m_\chi} \int d^3v \, v f({\bf v}) \frac{d \sigma}{dq^2}(q^2, v) \;,
\eea
where the differential SI WIMP-nucleus scattering cross-section is given by
\bea
	\frac{d \sigma}{dq^2}(q^2, v) &= \frac{\left[ Z f_p + \left( A-Z \right) f_n \right]^2}{\pi v^2} F^2(q) \Theta(q_{\rm max} - q)
\nonumber	\\ &\approx \frac{1}{4 \mu_p^2 v^2} A^2 \sigma_{\rm SI}  F^2(q) \Theta(q_{\rm max}-q) \;,\label{eq:diffXsec_approx}
\eea
with $Z$ as the atomic number, $A$ the atomic mass, $f_p$ the dark matter coupling to protons, $f_n$ the dark matter coupling to neutrons, and $F$ the nuclear form factor. The form factor accounts for the fact that as the energy of the dark matter particle increases, it will eventually only elastically scatter off individual nucleons rather than coherently scattering off the nucleus. We have assumed isospin conserving couplings (i.e. $f_p \simeq f_n$) in the second equivalence to write the recoil rate in terms of the SI WIMP-nucleon cross section, $\sigma_{\rm SI}$, and the reduced mass of the WIMP-nucleon system, $\mu_p$. Also, note that, for the small momentum transfers typical of low mass WIMPs scattering off nuclei, the nuclear form factor $F^2(q) \approx 1$. 

The Heaviside step function in Eq.~\ref{eq:diffXsec_approx}, $\Theta(q_{\rm max}-q)$, which arises from the maximal momentum transfer allowed by the elastic scattering kinematics, can be exchanged for a minimum velocity imposed on the phase space integral in Eq.~\ref{eq:diffRecoilRate}, which we can rewrite as
 \bea \label{eq:diffRecoilRate2}
		\frac{dR}{dE_R} = \frac{ \rho_{ \chi}}{m_\chi} \frac{ A^2}{2 \mu_p^2} \sigma_{\rm SI}  \int_{v > v_{\rm min}} d^3v \, \frac{ f({\bf v})}{v} \;,
\eea    
where $v_{\rm min} = \sqrt{M E_R / 2 \mu^2}$ for a target nucleus of mass $M$ with the reduced mass of the WIMP-nucleus system $\mu$. We assume the WIMP velocity distribution is given by a truncated Maxwell-Boltzmann distribution of the form
\bea  \label{eq:vdist}
	f({\bf v}) = \frac{1}{N_{\rm esc} \left( \pi v_0^2 \right)^{3/2}} \exp\left( - \frac{\left| {\bf v} + {\bf v}_E \right|^2}{v_0^2 } \right) \Theta (v_{\rm esc} - \left| {\bf v} + {\bf v}_E \right|) \;,
\eea
where ${\bf v}_E$ is the velocity of the Earth relative to the galactic rest frame, $v_0$ is the local circular velocity and $v_{\rm esc}$ is the local escape velocity. The normalization of the velocity distribution is given by    
\bea
	N_{\rm esc} = {\rm erf}\left(\frac{v_{\rm esc}}{v_0} \right) - \frac{2}{\sqrt{\pi}} \frac{v_{\rm esc}}{v_0} \exp\left(-\frac{v_{\rm esc}^2}{v_0^2} \right) \;.
\eea
In a self-consistent model of the Milky Way, the astrophysical parameters $\rho_\chi$, $v_0$ and  $v_{\rm esc}$ will not be independent~\cite{Benito:2016kyp,Green:2017odb,Petac:2018wnu,Lavalle:2014rsa}. However, given both the variation between different observations and the large uncertainties of individual measurements, for simplicity we ignore correlations between astrophysical parameters in the SHM.

We note that, for low mass WIMPs such that $m_\chi \ll M$, we have $v_{\rm min} \simeq \sqrt{M E_R / 2 m_\chi^2}$. Thus, for a fixed nuclear recoil energy $E_R$, the minimum kinematically allowed WIMP velocity will increase for smaller WIMP masses. Also, the mean inverse speed, given by the integral over the velocity distribution in Eq.~\ref{eq:diffRecoilRate2}, becomes increasingly dependent on the value of $v_{\rm esc}$ as $m_\chi$ decreases. While the value for $v_{\rm esc}$ in Eq.~\ref{eq:vdist} determines the lower threshold for sensitivity to WIMPs with smaller $m_\chi$ at a given experiment, the sensitivity to WIMPs with masses slightly above this threshold will also be determined by the shape of the velocity distribution, characterized by the local circular velocity $v_0$. In contrast, $v_{\rm min}$ is low enough for high mass WIMPs that the sensitivity of DD experiments is only slightly dependent on the shape and cut-off of the velocity distribution.

\subsection{Local dark matter density}
In DD experiments, the most commonly used values for the local dark matter density are either $\rho_{\chi}=0.3\, \mathrm{GeV/cm^3}$ or $\rho_{\chi}=0.4\, \mathrm{GeV/cm^3}$, while a considerably wider range of measured values is possible. Recent measurements suggest the local dark matter density should lie in the range of $(0.2-0.6)\, \mathrm{GeV/cm^3}$ as discussed in Refs.~\cite{Green:2017odb,Read:2014qva}. The relative errors of individual measurements of the local dark matter density are dominated by systematics and are typically $\approx$ 30\% of the central value, as suggested by Ref.~\cite{Evans:2018bqy}. In our work, we present the effects of variation in the local dark matter density on the sensitivity of DD experiments in two ways. First we sample a uniform distribution of $\rho_{\chi}$ between $(0.2-0.6)\,\mathrm{GeV/cm^3}$. As the individual error bars of each of the measurements of the dark matter density do not overlap for the most part, we sample a uniform distribution over the most likely range of the density.
Secondly, since the sensitivity of a given DD experiment has a simple linear scaling with $\rho_{\chi}$, we also wanted to independently analyze the effects of the uncertainty in the velocity distribution by choosing four discrete values of $\rho_{\chi}:$ $(0.3,\,0.4,\,0.5,\,0.6)\,\mathrm{GeV/cm^3}$. 

Finally, to check the impact of uncertainties from individual measurements of the dark matter density, we also sample a Gaussian distribution with central value $0.3\,\mathrm{GeV/cm^3}$ and standard deviation $0.09\,\mathrm{GeV/cm^3}$ (30\% of the central value). Assuming the Gaussian distributions for the local circular and escape velocities decribed below, we find that the sets of exclusion curves in the case of a Gaussian dark matter density distribution are essentially the same as exclusion limits in the case of the uniform distribution described above. The reason for this similarity is straightforward given the linear relationship between the exclusion limits and the dark matter density; the values of the dark matter density falling within $\sim 1\sigma$ of the central value corresponds to a range which resembles a uniform distribution of $(0.2-0.4)\,\mathrm{GeV/cm^3}$. Below we will therefore present results only for the two cases of uniform density distribution between $(0.2-0.6)\,\mathrm{GeV/cm^3}$ and the four discrete values listed in the previous paragraph.

\subsection{Local circular velocity}
For the value for the local circular velocity of the Sun around the center of galaxy we will use $v_0=220 \, \mathrm{km/s}$~\cite{Green:2017odb}. The largest uncertainty for the value of $v_0$ among recent measurements arises from the orbit of the GD-1 stellar stream, which has a value of 18 $\mathrm{km/s}$ or about 8.1\% of the central value. Since the error bar for $v_0$  in this case is symmetric about the central value, we assume $v_0$ will follow a normal distribution with central value $v_0= 220 \, \mathrm{km/s}$ and dispersion $18 \, \mathrm{km/s}$. We note that more precise determinations of the local circular velocity with central values of $\sim$230$\, \mathrm{km/s}$ have recently been reported in Refs.~\cite{Mroz2019,Eilers2019}, while similarly high values of the local circular velocity can be inferred from the proper motion of Sgr A*~\cite{Reid:2004rd} with recent estimates of the solar radius (for example, see~\cite{Abuter:2018drb,Abuter:2019}). Although assuming a central value of $v_0= 230 \, \mathrm{km/s}$ can have a small impact on our results near the low WIMP mass threshold for DD experiments, the differences between the most probable values of the exclusion curves are negligible compared to uncertainty bands, which span at least one order of magnitude in the SI WIMP-nucleon cross section at low WIMP masses. In addition, assuming a more precise measurment of the circular velocity with an uncertainty closer to 1\% of the central value causes the widths of the uncertainty bands to shrink by an ${\cal O}(1)$ factor which depends on $m_\chi$ and the other astrophysical parameters.

\subsection{Local escape velocity}
We compare the results of DD experiments using two measurements of the local escape velocity:
the latest result $v_{esc}=580\pm63\, \mathrm{km/s}$ from the DR2 of the Gaia survey~\cite{Monari:2018} (referred to as Monari $et\,al.$ escape velocity) and an older (commonly used) value  $v_{esc}=533^{+54}_{-41}\, \mathrm{km/s}$~\cite{Piffl:2013mla} from the RAVE survey (referred to as RAVE escape velocity). The errors of these two measurements are about 10\% of the central values. In the Monari $et \,al.$ result, the error bar is symmetric about central value, so for $v_{esc}$ we again assume it will follow a normal distribution. In the RAVE result, the uncertainties are asymmetric and, thus, applying a method suggested by Ref.~\cite{Barlow:2004wg}, the escape velocity from RAVE survey can be assumed to follow a probability distribution given by
\bea\label{eq:pdf}
	f(v_{esc})=\frac{1}{N}\exp\left(-\frac{1}{2}\left[\frac{\ln\left(1+\frac{v_{esc}-v_{0esc}}{\gamma}\right)}{\ln\beta}\right]^2\right) ,
\eea
where $v_{0esc}=533\, \mathrm{km/s}$ is the central value, $\gamma=\frac{\sigma_{+}\sigma_{-}}{\sigma_{+}-\sigma_{-}}$, $\beta=\frac{\sigma_{+}}{\sigma_{-}}$, $\sigma_{+}=54\, \mathrm{km/s}$, $\sigma_{-}=41\, \mathrm{km/s}$, and $N$ is the normalization factor. 
Note that in this choice for the distribution function, the probability that $v_{esc}\leq362\, \mathrm{km/s}$ is zero. However, $v_{esc}=362\, \mathrm{km/s}$ is significantly less than the central value of  $v_{0esc}=533\, \mathrm{km/s}$, so this assumption of distribution function does not affect our results.  One of the main purposes of this work is to determine how the different measurements of the escape velocity impact the exclusion limits of DD experiments. We note that a recent determination of the local escape velocity, which also uses the most recent Gaia data but does not assume the velcoity distribution of halo stars is isotropic, gives a value of  $v_{esc}=528^{+24}_{-25}\, \mathrm{km/s}$~\cite{Deason:2019}. Although the central value given in Ref.~\cite{Deason:2019} is more consistent with the RAVE escape velocity, we will demonstrate that even the larger central value of the Monari $et\,al.$ escape velocity does not broadly change the effects of astrophysical uncertainties on the senstivity of DD experiments.

\section{Results}
\label{sec:Results}
We have performed Monte Carlo simulations of the generated exclusion curves for XENON1T experiment based on their 2017 analysis~\cite{Aprile:2017iyp}. At a given dark matter mass, we draw $10^5$ points from the previously described distributions of the three astrophysical parameters and calculated the associated exclusion limits on the SI WIMP-nucleon scattering cross section using the DDCalc software~\cite{Workgroup:2017lvb}. The most probable value at a given WIMP mass is located at the peak of the histogram of the generated exclusion limits. In the following sections, the $2\sigma$ region around the respective peaks of the histograms for each WIMP mass is also referred to as the uncertainty band. We note that, due to the loss of detector efficiency and correspondingly high WIMP velocities required for low mass WIMPs to induce nuclear recoils above threshold, DDCalc only calculates limits based on the XENON1T analysis for WIMP masses $m_\chi \geq 3.9 \,$GeV. Since the sensitivity of DD experiments to elastic scattering induced by higher mass WIMPs only trivially depends on $\rho_\chi$, we also require $m_\chi \leq 15 \,$GeV in our analysis.     

\begin{figure}
\begin{center}
\includegraphics[width=10cm]{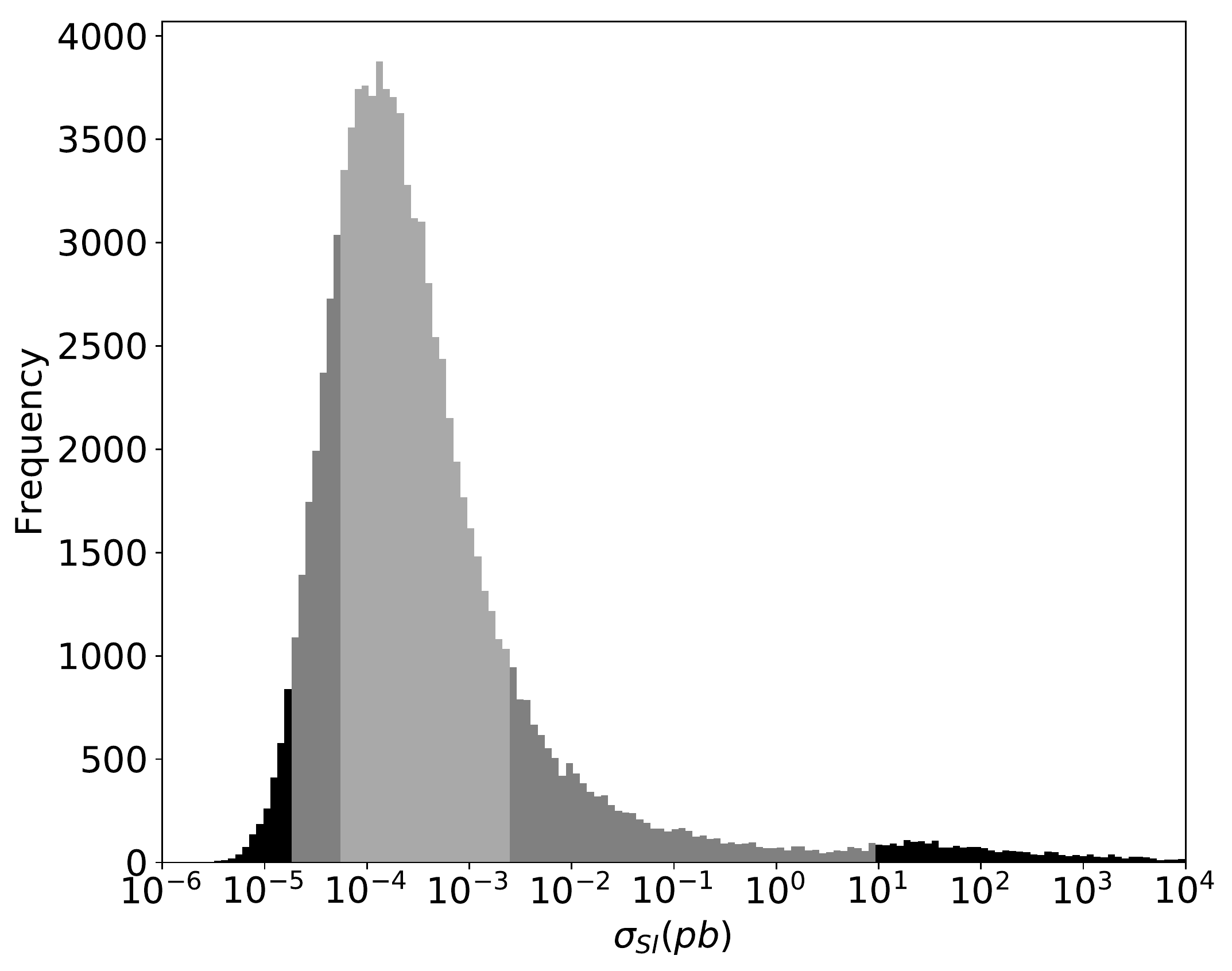}
\caption{A histogram of $10^5$ exclusion limits on SI WIMP-nucleon scattering cross section set by the XENON1T experiment~\cite{Aprile:2017iyp} for $m_\chi=4 \,$GeV and sampling over three astrophysical parameters: the local dark matter density $\rho_{\chi}$ which is assumed to be a uniform distribution between 0.2 and $0.6~\mathrm{GeV/cm^3}$, while the other two parameters are Gaussian distributed with local escape velocity $v_{esc}=580\pm63$~km/s (the Monari result) and local circular velocity $v_0=220\pm18$~km/s. The $1\sigma$ and $2\sigma$ regions are shaded light and dark grey, respectively, on the histogram, while the black shaded region lies outside of the $2\sigma$ uncertainty band. The corresponding histogram for the RAVE~\cite{Piffl:2013mla} escape velocity (not shown) looks very similar with small shift to weaker cross section limits. The uncertainty in the cross section limit at low WIMP masses is dominated by the escape velocity, causing the very long tail towards weaker limits due to the very strong suppression of the number of expected recoil events above threshold in a DD experiment.}
\label{fig:hist}
\end{center}
\end{figure}

An example histogram is shown using the Monari escape velocity for a 4~GeV WIMP in \figref{hist} with the $1\sigma$ region shaded light grey and the $2\sigma$ region shaded dark grey, while the black shaded region lies outside of the $2\sigma$ uncertainty band. We find that the distribution of limits on the SI WIMP-nucleon scattering cross section has a very long tail extending towards the higher/weaker cross section limits. The long tail corresponds to the smaller values of $v_{esc}$, where the sensitivity of an experiment is highly dependent on the escape velocity. In this region of parameter space with a small WIMP mass and small value for the escape velocity, there is a very strong suppression of the number of expected events above the experimental threshold for the SHM. Thus, a modest change in $v_{esc}$ can lead to an orders of magnitude variation in the exclusion limit. Because the relative uncertainties of the Monari and RAVE escape velocity measurements are similar and the central value of the RAVE result is lower than the central value in Monari $et\,al.$, the lower end of the RAVE $v_{esc}$ distribution reaches smaller velocities. Thus, one would expect not only a weakening of the most probable limits at low WIMP masses assuming the RAVE escape velocity, but also a slightly longer tail in the distribution of exclusion limits at higher cross sections. 

\subsection{Sampling three parameters}
In this subsection, we present the results of our Monte Carlo simulation of $10^5$ points drawn from a uniform distribution for $\rho_{\chi}$ between $(0.2-0.6)\,\mathrm{GeV/cm^3}$ as well as the previously described distributions for the local escape velocity and the local circular velocity. 
\begin{figure}
\begin{center}
\includegraphics[width=\linewidth]{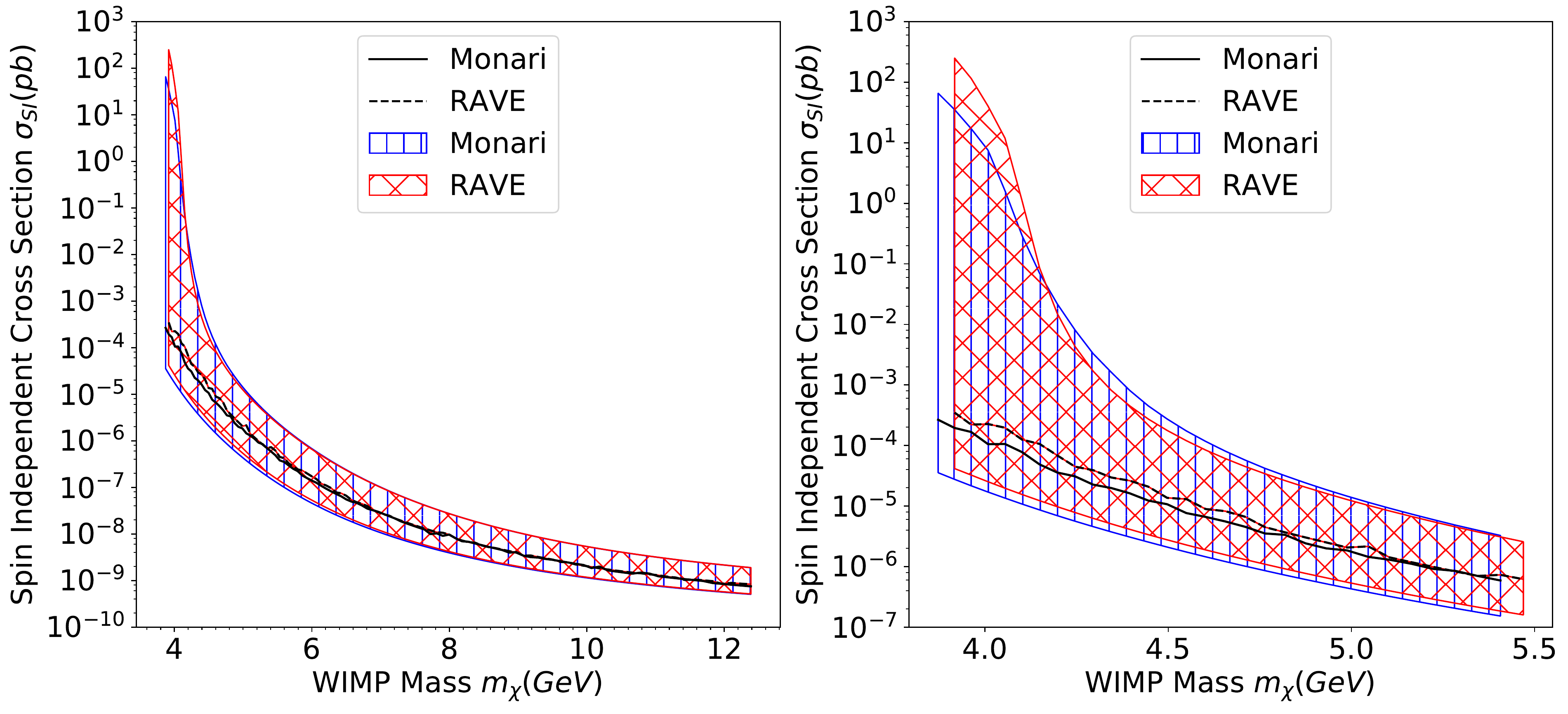}
\caption{The most probable value of the exclusion curve from the XENON1T experiment~\cite{Aprile:2017iyp} for the SI WIMP-nucleon cross section along with the corresponding $2\sigma$ ranges (hatched regions) using the Monari and RAVE escape velocities. A uniform distribution for $\rho_\chi$ between 0.2 and 0.6 $\mathrm{GeV/cm^3}$ is sampled while $v_0$ and $v_{esc}$ are distributed as described in the text. Left panel: exclusion curves in the range of WIMP masses 3.9--15~GeV. Right panel: exclusion curves zoomed in to WIMP masses below 6~GeV.}
\label{fig:3_parameter}
\end{center}
\end{figure}
In \figref{3_parameter}, the most probable values as well as the $2\sigma$ regions of exclusion curves for both the Monari escape velocity and the RAVE escape velocity are plotted based on the analysis of XENON1T data. Very little difference is found in the most probable exclusion curves assuming either the Monari or RAVE distributions for the escape velocity. Even when the WIMP mass is below 6~GeV where the difference is the greatest, the two curves are very similar and any deviations remain well within the $2\sigma$ ranges. As expected, the Monari $et \, al.$ most probable exclusion curve is slightly below the RAVE most probable exclusion curve because the Monari escape velocity has a larger central value for $v_{esc}$ than the RAVE result. The sensitivity of DD experiments to low mass WIMPs increases with larger escape velocities because the higher WIMP velocities induce nuclear recoils with energies sufficiently above the experimental threshold. Similarly, the smaller escape velocities from the RAVE distribution explains why the $2\sigma$ range extends to higher cross sections at low WIMP masses. 

\subsection{Sampling two parameters}
In this subsection, we fix the local dark matter density to four discrete values in the range $(0.2-0.6)\,\mathrm{GeV/cm^3}$ and draw $10^5$ Monte Carlo samples from the distributions for $v_0$ and $v_{esc}$.  The purpose of this sampling strategy is to determine the effects of the new, larger central value of the escape velocity from the Gaia data independently of variations in the dark matter density.
\begin{figure}[!h]
\begin{center}
\includegraphics[height=3 in]{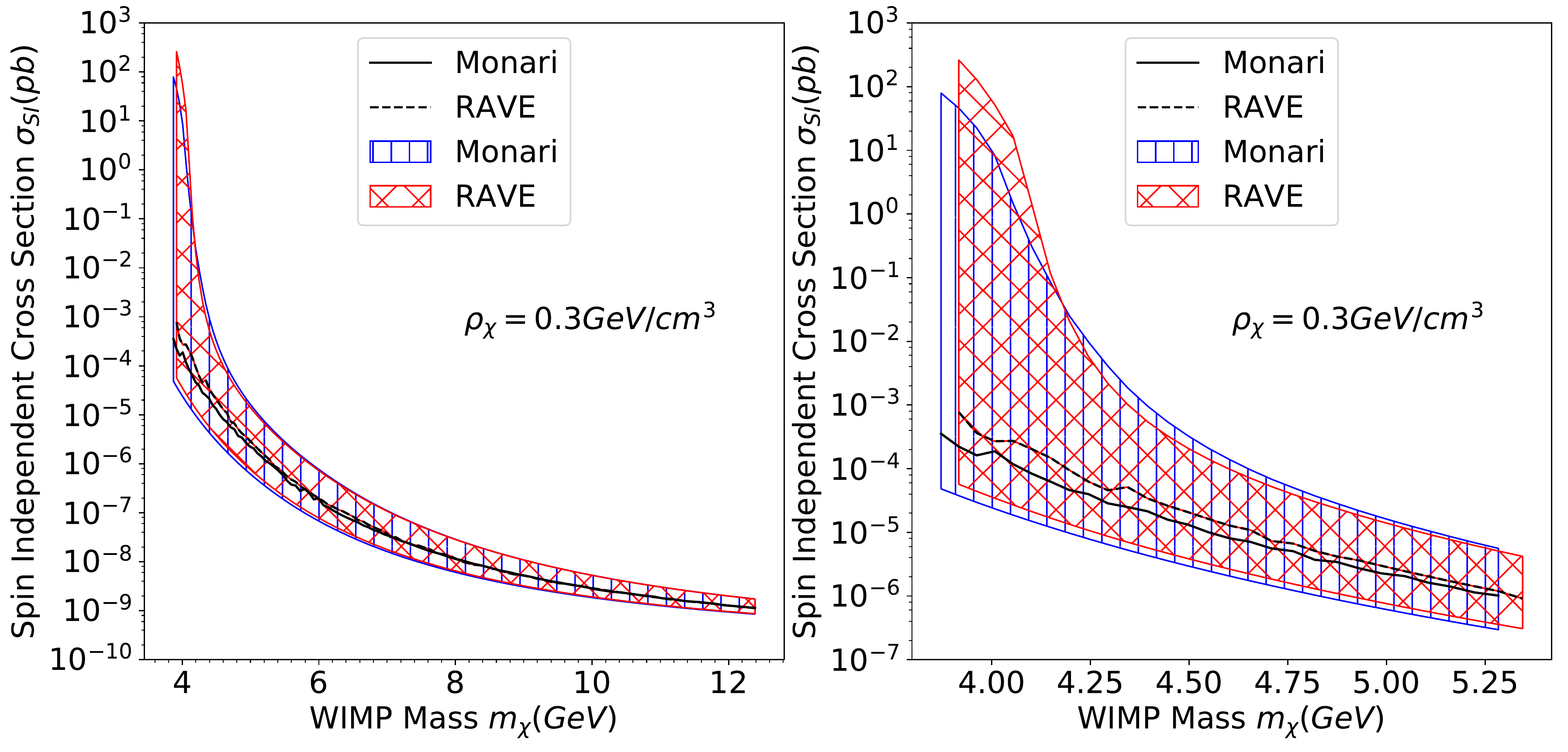}
\caption{The most probable value of the exclusion curve from the XENON1T experiment~\cite{Aprile:2017iyp} for the SI WIMP-nucleon cross section along with the corresponding $2\sigma$ ranges (hatched regions) using the Monari and RAVE escape velocities. Here $\rho_\chi$ is taken to be 0.3$\, \mathrm{GeV/cm^3}$  Left panel: exclusion curves in the range of WIMP masses 3.9--15~GeV. Right panel: exclusion curves zoomed in to WIMP masses below 6~GeV. Above WIMP mass 6~GeV, comparison to \figref{3_parameter} indicates the uncertainty from local dark matter density begins to dominate; below 6~GeV, the uncertainty from escape velocity dominates.} 
\label{fig:sample_two_variables}
\end{center}
\end{figure}

We show the most probable exclusion curves as well as the uncertainty bands corresponding to a fixed value of $\rho_\chi = 0.3\, \mathrm{GeV/cm^3}$ in \figref{sample_two_variables}. We find the uncertainty bands shrink quickly as the WIMP mass increases, from about 6 orders of magnitude at 4~GeV to a factor of $\sim$1.6 at 15~GeV. In the previous subsection, in which $\rho_\chi$ is uniformly distributed, the width of uncertainty bands becomes approximately constant at a factor of $\sim$3 as the WIMP mass approaches 15~GeV. From this comparison, we can conclude that the uncertainty of the exclusion curve at WIMP masses greater than 6~GeV mainly comes from the uncertainty in local dark matter density. A more accurate estimate of $\rho_\chi$ could significantly reduce the width of the uncertainty bands on the corresponding exclusion curves. We also notice the consistency of the most probable exclusion curves for Monari $et\, al.$ and RAVE when either scanning over or fixing the dark matter density. This consistency reflects the independence of exclusion limits for WIMPs with masses above 6~GeV from uncertainties in the best fit values of the escape velocity.

At lower WIMP masses, however, a difference between the most probable exclusion curves for Monari $et\, al.$ and RAVE can be observed, as displayed in \figref{sample_two_variables}.  Comparing \figref{3_parameter} and \figref{sample_two_variables}, we can see the differences between the most probable exclusion curves for limits calculated using the Monari $et\, al.$ and RAVE data found in \figref{sample_two_variables} are similar to differences between the most probable exclusion curves when scanning over the uncertainty in local dark matter density in \figref{3_parameter}. Even though the most probable exclusion curves can differ significantly at low WIMP masses, the uncertainty bands using the Monari $et\, al.$ and RAVE data broadly overlap regardless of whether the dark matter density is fixed or not. Thus, more precise measurements of the escape velocity could significantly reduce the astrophysical uncertainties associated with the sensitivity of DD experiments to low mass WIMPs.

We also performed Monte Carlo simulations of additional fixed values of the dark matter mass extending up to 0.6$~\mathrm{GeV/cm^3}$.  In each of these cases, the overall shapes of the curves remained approximately constant.  The main change is the curves are scaled vertically to lower/stronger values of the SI WIMP-nucleon cross section limit at all WIMP masses by the ratio of the dark matter density to 0.3$~\mathrm{GeV/cm^3}$. As the nuclear recoil spectrum is directly proportional to $\rho_\chi$, this is the expected scaling of cross section limits with the local dark matter density.

\section{Conclusions}
\label{sec:Conclusions}
In this work we have examined the effects of astrophysical uncertainties on the exclusion limits set by XENON1T~\cite{Aprile:2017iyp}.  While this work was originally motivated by the new value for the escape velocity measured using Gaia data \cite{Monari:2018}, we expanded our analysis to include the uncertainties on the local dark matter density and speed of the Sun's orbit around the galactic center. We utilized the DDCalc software \cite{Workgroup:2017lvb} to perform the likelihood analyses.  To our knowledge, this is the first work that uses DDCalc to quantify these effects specifically for the standard halo model.  As this is the model that is canonically assumed in the production of all direct detection exclusion limits, understanding the impact of the astrophysical uncertainties of this model  on these limits is important.  These uncertainties are typically neglected when collaborations present their final limits.

Several general features emerge from our analysis. Above a WIMP mass of 6~GeV, the uncertainty of the limits on the SI WIMP-nucleon cross section is dominated by the local dark matter density. The $2\sigma$ range for the exclusion limit when sampling a uniform distribution of $\rho_{\chi}$ approaches a factor of $\sim$3 as the WIMP mass increases to 15~GeV and a more accurate estimate on $\rho_\chi$ is thus the most important parameter to reduce the effect of astrophysical uncertainties in this mass range. At smaller WIMP masses, where the uncertainty of the escape velocity begins to dominate, we have demonstrated that the associated uncertainty band is $\sim$6 orders of magnitude wide in the cross section near a WIMP mass of 4~GeV. This band is highly asymmetric extending to larger cross sections (weaker limits) as the WIMP mass decreases. The experimental sensitivity is a very strong function of $v_{esc}$ as the WIMP mass decreases, leading to much weaker exclusion limits for fairly modest decreases in the escape velocity.

The astrophysical uncertainties calculated in this work are larger than the statistical fluctuations in the background presented in the XENON1T cross section limits at low WIMP masses. For example, the $2\sigma$ band of the expected limit from the statistical fluctuations in the background at $m_\chi = 7 \,$GeV is less than a factor of $\sim$3 in the XENON1T analysis. By contrast, when we assume a fixed $\rho_{\chi}$ while allowing $v_0$ and $v_{esc}$ to vary given their respective uncertainties in \figref{sample_two_variables}, the astrophysical uncertainties yield more than a factor of $\sim$6 variation in the exclusion limit on the SI WIMP-nucleon scattering cross section for $m_\chi = 7 \,$GeV. If properly incoperated into the Xenon1T likelihood analysis, astrophysical uncertainties in the signal can weaken exclusion limits on the SI WIMP-nucleon scattering cross section at low WIMP masses while the statistical fluctuations in the background should remain roughly constant. The effects of astrophysical uncertainties are important to consider when comparing limits generated by xenon-based experiments to experiments focused on detecting low mass dark matter. Targets in low mass dark matter searches are often composed of lighter nuclei and the experiments typically have lower nuclear recoil energy thresholds. Although the astrophysical uncertainties from the local escape velocity will still eventually dominate the uncertainty, results from direct detection experiments with lighter target nuclei and lower thresholds should be more robust at lighter WIMP masses.       

Additionally, we find that the updated best fit value for the escape velocity based on Gaia data does not have a large effect on the astrophysical uncertainties related to the XENON1T limit. In particular, the differences between the most probable exclusion limits for the different measurements of the escape velocities are well within the 2$\sigma$ uncertainty ranges for the astrophysical uncertainties.

\acknowledgments
The authors would like to thank S.~Baum, P.~F.~de Salas and K.~Hattori for useful discussions. YW, KF, and PS acknowledge support from DoE grant DE-SC007859 and the LCTP at the University of Michigan. KF and PS acknowledge support by the Vetenskapsr\r{a}det (Swedish Research Council) through contract No. 638-2013-8993 and the Oskar Klein Centre for Cosmoparticle Physics. MV acknowledges support from NSF award AST-1515001 and NASA-ATP award NNX15AK79G.

\end{document}